\documentclass[conference]{IEEEtran}
\usepackage[T1]{fontenc}
\usepackage{helvet} %
\usepackage{cite}
\usepackage{amsmath}
\DeclareMathVersion{sans}
\SetSymbolFont{operators}{sans}{OT1}{phv}{m}{n}
\SetSymbolFont{letters}{sans}{OML}{cmbrm}{m}{it}
\SetSymbolFont{symbols}{sans}{OMS}{cmbrs}{m}{n}
\usepackage{booktabs}
\usepackage{graphicx}
\usepackage{stfloats}
\usepackage{xcolor}
\usepackage{tikz}
\usepackage{pgfplots}
\pgfplotsset{compat=1.18}
\usetikzlibrary{arrows.meta,positioning}
\usepackage[hidelinks]{hyperref}
\title{LLM Agent-Based Design of RISC-V CPUs\\
with FPGA-in-the-Loop Evaluation}
\author{
\IEEEauthorblockN{Leo Itoh}
\IEEEauthorblockA{\textit{The University of Tokyo}\\
Tokyo, Japan\\
itoh@mtl.t.u-tokyo.ac.jp}
\and
\IEEEauthorblockN{Junichiro Kadomoto}
\IEEEauthorblockA{\textit{The University of Tokyo}\\
Tokyo, Japan\\
kadomoto@mtl.t.u-tokyo.ac.jp}}
\begin{document}
\AddToHookNext{shipout/foreground}{\put(0.5\paperwidth,-0.45in){\makebox[0pt]{\parbox{0.8\paperwidth}{\centering\footnotesize This work has been submitted to the IEEE for possible publication. Copyright may be transferred without notice, after which this version may no longer be accessible.}}}}

\maketitle
\begin{abstract}
Long-run RTL simulation of CPU candidates can take hours, while short diagnostic runs may miss performance degradation that emerges later in execution.
This evaluation cost can delay design decisions in LLM-agent-based CPU development.
We present a development environment in which LLM agents implement CPU candidates on an FPGA and use on-board performance measurements to guide design decisions.
We applied it to the human-supervised development of a single-cycle RV32I core into a three-wide out-of-order RV32IM core.
The final design reaches 5.61 CoreMark/MHz at 130 MHz on a Artix-7 FPGA.
For the same CPU candidate running 8,000 CoreMark iterations, FPGA evaluation, including implementation, was 9.6$\times$ faster than single-threaded RTL simulation.
In one case, board measurements revealed long-run performance degradation in a candidate that improved in a short diagnostic run.
The result guided candidate selection and led to RTL-based root-cause analysis and a fix.
\end{abstract}

\section{Introduction}
\label{sec:introduction}
\begin{figure*}[t]
\centering
\includegraphics[width=\textwidth]{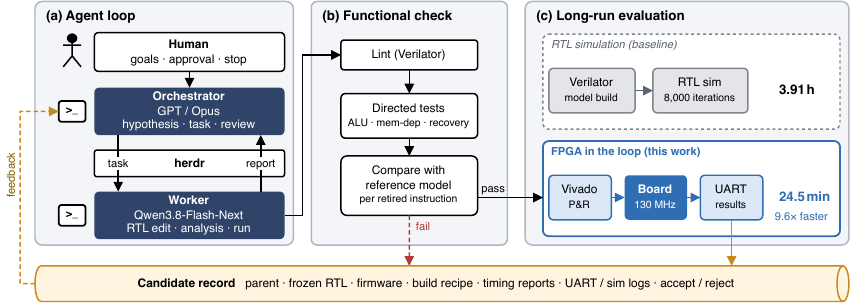}
\caption{Agent-based CPU development environment.
Functional checks and on-board performance measurements feed back into agent decisions through a shared candidate record.}
\label{fig:framework}
\end{figure*}
The performance and implementation cost of a CPU core depend on the choice of pipeline, issue width, branch prediction, and memory system. The open RISC-V instruction set architecture has made it practical to design cores that explore these choices.
Kadomoto et al. designed RISC-V cores with different pipeline structures and compared their performance, power, and area~\cite{kadomoto2022}.
Fu et al. evaluated single-issue, multi-issue, and out-of-order cores with a common SoC and implementation flow~\cite{fu2025}.
Beyond such comparisons, BOOM-Explorer searches BOOM microarchitecture parameters by multi-objective Bayesian optimization with active learning and a surrogate model~\cite{bai2023}.
ChatDSE explores a parameterized design space using LLM-based estimates of performance and power~\cite{tang2025}.

LLMs also generate and modify RTL.
ChatCPU supports CPU design and verification from a specification and reported a RISC-V CPU tape-out~\cite{wang2024}.
El-Helwani and Khalid compared prompt structures for generating a single-cycle RV32I core and evaluated its FPGA implementation~\cite{elhelwani2026}.
BluesFL maps instruction execution to code structure to support LLM-based fault localization in CPUs~\cite{liu2026}.

However, full-length RTL simulation can take hours to evaluate a CPU candidate.
Short diagnostics reduce evaluation time but may miss performance degradation caused by predictor and scheduler state that develops during longer runs.

Prior work has also integrated FPGA implementation into agent-based CPU development.
In auto-arch-tournament, agents modify CPU RTL and perform functional checks, followed by FPGA place and route~\cite{bonetto_hwebench_2026}.
Its standard evaluation code runs CoreMark on Verilator and multiplies iterations per cycle by the Fmax from place and route.
Programming and running a physical board are not part of this evaluation path.

FPGA execution also supports cycle-exact system simulation in FireSim~\cite{karandikar2018} and DNN accelerator design space exploration in Quickloop~\cite{inayat2024}.
Lyra uses instruction streams from a generative model for differential testing and coverage collection on RISC-V FPGA cores~\cite{huo2026}.
Our focus is on feeding long-run board measurements back to LLM agents for CPU candidate selection and RTL analysis.

We make three contributions.
First, we present a development environment that connects agent-driven RTL changes to on-board evaluation under human supervision.
Second, we report the development of an RV32I core into a three-wide out-of-order RV32IM core that exceeds published open-source 32-bit soft cores in CoreMark/MHz, with long-run board results guiding selection and analysis.
Third, we compare the time to evaluate the same candidate by RTL simulation and on an FPGA.
\section{Development Framework}
\label{sec:framework}
\begin{figure}[!t]
\centering
\includegraphics[width=.70\columnwidth]{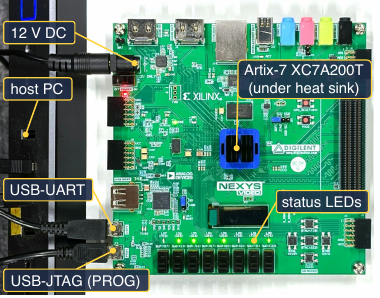}\\[1pt]
{\footnotesize (a)}\\[3pt]
\includegraphics[width=.88\columnwidth]{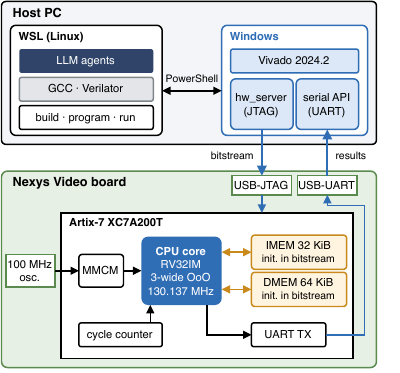}\\[1pt]
{\footnotesize (b)}
\caption{FPGA testbed: (a) Nexys Video board and (b) host connection and on-chip design.}
\label{fig:testbed}
\end{figure}
Agents extend a single-cycle RV32I core using the verification and FPGA evaluation flow in Fig.~\ref{fig:framework}.

\subsection{Human and Agent Roles}
The human set goals at each stage of CPU development and proposed directions such as multiple issue and out-of-order execution.
The human also approved work and decided when to stop.
Agents wrote and modified the CPU RTL and carried out detailed design, implementation, optimization, analysis, and verification.
We used both single-agent and orchestrator--worker configurations during development.
In the latter, GPT models~\cite{openai_gpt61sol} or Claude Opus 5.5~\cite{anthropic_opus55} coordinated design and review, while the Qwen3.8-Flash-Next model~\cite{qwen2026design} handled implementation and analysis.
Tasks and reviews were exchanged through herdr v0.9.3~\cite{herdr}.

The orchestrator issues tasks that specify the change hypothesis, target files, verification conditions, acceptance criteria, and compute allocation.
Each worker implements changes and evaluates candidates in a separate Git worktree, then reports the results.

\subsection{FPGA Evaluation}
Fig.~\ref{fig:testbed} shows the evaluation setup.
Agents in WSL2 call Vivado, hw\_server, and the serial API on the Windows side.
Resource usage and timing come from Vivado reports, and cycle counts and CRCs come from the board UART.

Functional verification uses lint, directed tests for arithmetic, memory dependence, and branch recovery, and a per-retirement comparison with a reference model.
In these tests, each retired instruction’s PC and register writeback are checked against a custom RV32IM instruction-set model implemented in Python.
Board measurements informed performance-based candidate selection before full-length RTL simulation completed.
Final acceptance still required matching UART output from a full-length simulation and two independent board runs using the frozen RTL and firmware.

\subsection{Experiment Management}
\begin{table}[t]
\caption{Recorded Development Activity}
\label{tab:loop}
\centering\footnotesize
\setlength{\tabcolsep}{3pt}
\begin{tabular}{@{}lp{.6\columnwidth}@{}}
\toprule
Activity & Observed count \\
\midrule
Human $\rightarrow$ orchestrator & 11 messages \\
Orchestrator $\rightarrow$ worker & 18 tasks, 24 reviews, 10 continuations or resumptions \\
Worker evaluations & 115 RTL simulations, 40 FPGA implementations, 4 board runs \\
\midrule
RTL candidates & 29 total; 27 simulated, 6 implemented on FPGA, 2 evaluated on board \\
\bottomrule
\end{tabular}
\par\smallskip
\parbox{\columnwidth}{Observed records for the three adopted changes, not a complete census. FPGA implementations include checkpoint resumptions.}
\end{table}%
Each CPU design candidate has an identifier and a link to its parent design.
We save the frozen RTL, firmware, implementation settings, logs, hashes, and the accept or reject decision.
Tasks and reviews include the latest measurement results and failure reasons.
Handover notes record measurements, running jobs, and next steps.
Table~\ref{tab:loop} summarizes the recorded development activity for three changes incorporated into the final CPU: a load-wait predictor fix, register-move aliasing, and a two-way BTB.
\section{Experimental Setup}
\label{sec:setup}
\begin{figure}[!t]
\centering
\includegraphics[width=\columnwidth]{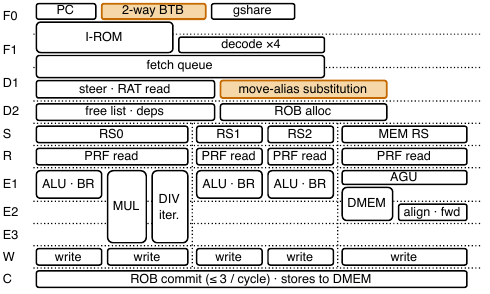}
\caption{Final CPU pipeline. Orange blocks highlight register-move aliasing and the two-way BTB.}
\label{fig:pipeline}
\end{figure}

\begin{figure}[!t]
\centering
\includegraphics[width=\columnwidth]{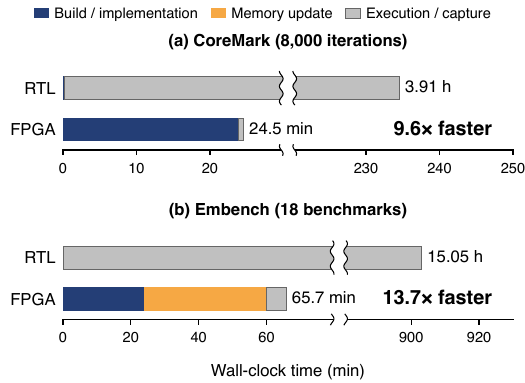}
\caption{Evaluation time including FPGA implementation. Axes are broken and scaled separately for each workload.}
\label{fig:time}
\end{figure}

\begin{figure*}[!t]
\centering
\includegraphics[width=\textwidth]{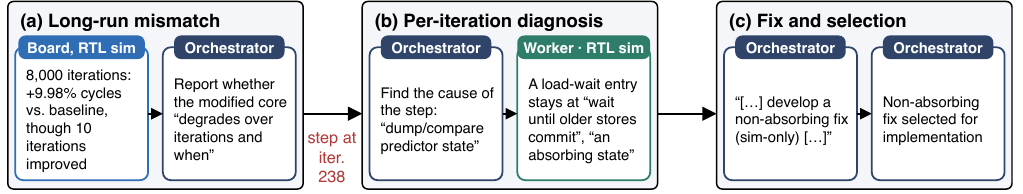}
\caption{Agent feedback in the predictor case study: (a) long-run evaluation, (b) RTL diagnosis, and (c) fix selection.
Quoted text is excerpted from the records ([\ldots] marks omissions). Unquoted text summarizes the records.}
\label{fig:loop}
\end{figure*}

\begin{figure}[!t]
\centering
\includegraphics[width=\columnwidth]{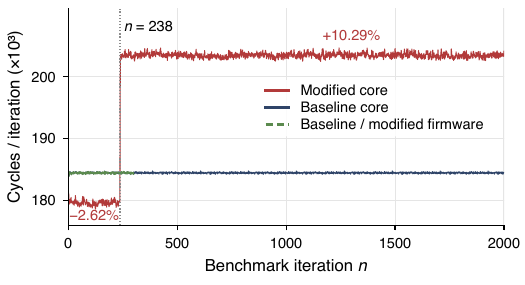}
\caption{Per-iteration RTL measurements for the predictor case study. The modified core enters a persistent slower state at iteration 238.}
\label{fig:rank-reversal}
\end{figure}

We run CoreMark for 8,000 iterations, corresponding to a measured interval of 10.95~s and satisfying its 10~s minimum runtime~\cite{coremark}.
We also evaluate 18 of the 19 Embench-IoT benchmarks, since \texttt{cubic} does not fit in the 32~KiB instruction memory.
The final design targets a bare-metal RV32IM environment.
It has a 32~KiB instruction memory and a 64~KiB data memory.
The program and initial data are stored as FPGA memory initialization values.

CoreMark performance is reported in iterations per second (iter/s).
Elapsed time is the board-reported tick count divided by the MMCM frequency.
IPC is the number of instructions retired during the measured interval divided by its tick count.
We obtain both counts from a simulation of the same RTL and firmware.

For the controlled CoreMark comparison, we use one frozen RTL and firmware configuration of the final design.
The firmware was built with RISC-V GCC 13.2.0 at -O3 with additional unrolling (\texttt{-funroll-loops} and related options) for 8,000 iterations.
The RTL simulation model was built with Verilator 5.020 and run on a single thread with assertions disabled.
All timings were measured on one host with an AMD Ryzen 7 5700X (8 cores, 16 threads) and 64~GB of memory, using Ubuntu 24.04 on WSL.
Verilator and GCC ran on Ubuntu, and Vivado ran on Windows.
We measured the wall-clock duration of each evaluation step.
Model build, full-length simulation, FPGA re-implementation, and result capture were each measured once.
Programming and result capture with an existing bitstream were measured three times.
FPGA evaluation includes implementation, memory initialization updates, programming, and result capture.
For Embench, one FPGA implementation is shared across all 18 benchmarks, with per-benchmark memory-initialization updates instead of repeated placement and routing.
RTL evaluation includes model build and full-length execution.
We timed successful evaluation runs after the shared firmware build.
The FPGA re-implementation reused previously established settings and met setup and hold timing on the first attempt.
Embench uses \texttt{CPU\_MHZ=130}.
\section{Results}
\label{sec:results}
\subsection{CPU Performance}
Table~\ref{tab:perf} reports CoreMark/MHz for representative development stages and existing soft cores.
Starting from the single-cycle RV32I core, we added a five-stage RV32IM pipeline, dual issue, and out-of-order execution with register renaming.
We then restructured the design to increase its implementation frequency.

\begin{table}[t]
\caption{CoreMark Performance}
\label{tab:perf}
\centering\footnotesize
\setlength{\tabcolsep}{3pt}
\begin{tabular}{@{}lrrr@{}}
\toprule
Design & CoreMark/MHz & MHz & LUT \\
\midrule
\multicolumn{4}{@{}l}{\textit{This work (on board)}} \\
\quad Single-cycle RV32I & 1.36 & 25 & -- \\
\quad 1-wide, 5-stage & 2.87 & 100 & -- \\
\quad 2-wide in-order & 3.32 & 100 & -- \\
\quad 3-wide OoO & 5.44 & 100 & -- \\
\quad 3-wide OoO (final) & 5.61 & 130 & 30,344 \\
\multicolumn{4}{@{}l}{\textit{Existing soft cores (published)}} \\
\quad VexRiscv full max perf~\cite{vexriscv_performance} & 2.57 & 200 & 1,935 \\
\quad NaxRiscv RV32~\cite{naxriscv_performance} & 5.02 & 155 & $\approx$13,300 \\
\bottomrule
\end{tabular}
\par\smallskip
\parbox{\columnwidth}{Published soft-core results use speed grade -3 (this work uses -1) and different memory configurations, compiler settings, and design scopes.}
\end{table}

The final design is a three-wide out-of-order RV32IM core with register renaming (Fig.~\ref{fig:pipeline}).
It combines timing-aware pipeline partitioning with register-move aliasing and a two-way BTB.
It reaches 730.6 iter/s at 130.137~MHz over 8,000 CoreMark iterations.
Its 5.61 CoreMark/MHz is the highest in Table~\ref{tab:perf}, 12\% above the published NaxRiscv result.
The design uses 10,320 FFs, 32 BRAM tiles, and 4 DSPs, with setup/hold slack of +0.003/+0.035~ns and an IPC of 1.60.
On Embench-IoT 1.0~\cite{embench}, it achieves a geometric-mean frequency-normalized score of 1.759 relative to the reference Cortex-M4 over 18 benchmarks.

\subsection{Evaluation Time}
Fig.~\ref{fig:time} compares RTL and FPGA evaluation times using identical RTL and firmware.
For CoreMark, RTL evaluation took 3.91~h and FPGA evaluation took 24.5~min, a speedup of 9.6$\times$.
For Embench, RTL evaluation took 15.05~h and FPGA evaluation took 65.7~min, a speedup of 13.7$\times$.
With an existing bitstream, programming and result capture for CoreMark take a median of 27.5~s.

\subsection{Case Study}
\label{sec:rank-feedback}
We evaluated a predictor-modified core with TAGE-lite and a return address stack.
It reduced cycles in a short 10-iteration diagnostic but increased the cycle count by 9.98\% over 8,000 iterations.
The baseline is the core before this extension, an intermediate design preceding the final core in Table~\ref{tab:perf}.
The implementation agent used two independent board runs to reject the predictor-modified core and retain the baseline.
Full-length RTL simulation reproduced the board tick count more than three hours later.

The orchestrator then assigned the mismatch to a worker for per-iteration RTL analysis (Fig.~\ref{fig:loop}(a)).
Fig.~\ref{fig:rank-reversal} shows the transition.
The modified core used 2.62\% fewer cycles over iterations 2--237, then entered a persistent slower state at iteration 238.
Over iterations 239--2,000, it used 10.29\% more cycles.

State tracing and ablation requested by the orchestrator identified an absorbing state in the load-wait predictor (Fig.~\ref{fig:loop}(b)).
A load replay writes a distance-7 prediction that makes the load wait for older stores to commit.
This prevents further replays and retraining while keeping a memory reservation station occupied.
Clearing the predictor entry instead of storing distance 7 eliminated the abrupt slowdown in a 400-iteration RTL run of the modified core.
The baseline running the modified-core firmware remains flat through 300 iterations, ruling out the firmware difference as the cause of the transition.
The orchestrator requested a non-absorbing fix.
After reviewing the worker's comparison, it selected a fix with no additional state (Fig.~\ref{fig:loop}(c)).

Vivado timing results also guided RTL changes.
The register-move source substitution had a setup slack of $-0.291$~ns at 130.137~MHz.
The orchestrator requested separating the substitution data from the instruction queue.
The revised version uses a separate LUTRAM and a ready bit.
The final design combines this revision with the two-way BTB and passed on-board acceptance.
\section{Conclusion}
\label{sec:discussion}
We presented an FPGA-in-the-loop environment for human-supervised, agent-based CPU development.
In the development case, on-board measurements guided candidate selection and subsequent RTL analysis and fixes.
Including the cost of the initial FPGA implementation, FPGA evaluation was 9.6$\times$ faster than RTL simulation for CoreMark and 13.7$\times$ faster for the 18 Embench benchmarks.
The resulting core reached 5.61 CoreMark/MHz at 130~MHz on board, above published VexRiscv and NaxRiscv results.
Selected development conversations from the long-run case study and representative CPU RTL snapshots are available online.\footnote{\url{https://github.com/kadomoto-lab/llm-cpu-fpga-loop}}

\section*{Acknowledgment}
We used OpenAI Codex and Anthropic Claude Code to assist with drafting and figure preparation, with all content reviewed and revised by the authors. This work was supported in part by JST CRONOS Grant Number JPMJCS25K2 and JST ASPIRE Grant Number JPMJAP2515.

\clearpage
\bibliographystyle{IEEEtran}
\bibliography{references}
\end{document}